\documentclass[a4,10pt]{article}

\usepackage{amssymb}
\usepackage{graphicx}
\usepackage[a4paper]{geometry}

\usepackage{moreverb}
\usepackage[utf8]{inputenc}

\newcommand {\apgt} {\ {\raise-.5ex\hbox{$\buildrel>\over\sim$}}\ }
\newcommand {\aplt} {\ {\raise-.5ex\hbox{$\buildrel<\over\sim$}}\ }

\title{MPWide: a light-weight library for efficient message passing over wide area networks}

\author{Derek Groen$^{1}$, Steven Rieder$^{2,3,4}$, Simon Portegies Zwart$^{2}$\\
\small{$^1$ Centre for Computational Science, University College London, London, United Kingdom}\\
\small{$^2$ Leiden Observatory, Leiden University, Leiden, The Netherlands}\\
\small{$^3$ System and Network Engineering research group, University of Amsterdam, Amsterdam, the Netherlands}\\
\small{$^4$ Kapteyn Instituut, Rijksuniversiteit Groningen, Groningen, the Netherlands}\\
\small{E-mail: d.groen@ucl.ac.uk}}
 
\date{\today}

\begin{document}
\maketitle

\begin{abstract}
We present MPWide, a light weight communication library which allows efficient message passing
over a distributed network. MPWide has been designed to connect application running on distributed
(super)computing resources, and to maximize the communication performance on wide area networks
for those without administrative privileges. It can be used to provide message-passing between 
application, move files, and make very fast connections in client-server environments. MPWide has 
already been applied to enable distributed cosmological simulations across up to four supercomputers 
on two continents, and to couple two different bloodflow simulations to form a multiscale simulation.
\end{abstract}


\section{Overview}

\subsection{Introduction}

Modern scientific software is often complex, and consists of a range of
hand-picked components which are combined to address a pressing scientific or
engineering challenge. Traditionally, these components are combined locally to
form a single framework, or used one after another at various locations to form
a scientific workflow, like in AMUSE~\footnote{AMUSE -
http://www.amusecode.org}~\cite{PortegiesZwart:2013,muse}. However,
these two approaches are not universally applicable, as some scientifically
important functionalities require the use of components which run concurrently,
but which cannot be placed on the same computational resource. Here we present
MPWide, a library specifically developed to facilitate wide area communications
for these {\em distributed applications}.

The main use of MPWide is to flexibly manage and configure wide area
connections between concurrently running applications, and to facilitate
high-performance message passing over these connections. These functionalities
are provided to application users and developers, as MPWide can be
installed and used without administrative privileges on the
(super)computing resources. We initially reported on MPWide in 2010~\cite{MPWide}, 
but have since extended the library considerably, making it more configurable 
and usable for a wider range of applications and users. Here we
describe MPWide, its implementation and architecture, requirements and reuse potential.

MPWide was originally developed as a supporting communication library in the
CosmoGrid project~\cite{CosmoGrid}. Within this project we constructed and
executed large cosmological N-body simulations across a heterogeneous global
network of supercomputers. The complexity of the underlying supercomputing and
network architectures, as well as the high communication performance required
for the CosmoGrid project, required us to develop a library that was both highly
configurable and trivial to install, regardless of the underlying
(super)computing platform.

There are a number of tools which have similarities to MPWide.
ZeroMQ~\cite{zeromq}, is a socket library which supports a wide range of
platforms.  However, compared with MPWide it does have a heavier dependency
footprint. Among other things it depends on {\tt uuid-dev}, a package that
requires administrative privileges to install. In addition, there are several
performance optimization parameters which can be tweaked with MPWide but not
with ZeroMQ. Additionally, the NetIBIS~\cite{NetIbis} and the
PadicoTM~\cite{PadicoTM} tools provide functionalities similar to MPWide,
though NetIBIS is written in Java, which is not widely supported on the compute 
nodes of supercomputers, and PadicoTM requires the presence of a central rendez-vous
server. For fast file transfers, alternatives include GridFTP and various
closed-source file transfer software solutions. There are also dedicated tools
for running MPI applications across clusters~\cite{Hockney,Manos,Agullo:2011}
and for coupling applications to form a multiscale simulation (e.g.,
MUSCLE~\cite{Borgdorff:2013} and the Jungle Computing
System~\cite{Drost:2012}).

\subsection{Summary of research using MPWide}

MPWide has been applied to support several research and technical projects 
so far. In this section we summarize these projects, the purpose for which MPWide
has been used in these projects, and the performance that we obtained using MPWide.

\subsubsection{The CosmoGrid project}

MPWide has been used extensively in the CosmoGrid project, for which it was
originally developed. In this project we required a library that enabled fast
message passing between supercomputers and which was trivial to
install on PCs, clusters, little Endian Cray-XT4 supercomputers and big Endian IBM
Power6 supercomputers. In addition, we needed MPWide to deliver solid communication
performance over light paths and dedicated 10Gbps networks, even when these networks
were not optimally configured by administrators.

In CosmoGrid we ran large cosmological simulations, and at times in parallel across
multiple supercomputers, to investigate key properties of small dark matter
haloes~\cite{cosmogridscience}. We used the GreeM cosmological N-body
code~\cite{Ishiyama09}, which in turn relied on MPWide to facilitate the fast 
message-passing over wide area networks. 

Our initial production simulation was run distributed,
using a supercomputer at SurfSARA in Amsterdam, and one at the National
Astronomical Observatory of Japan in Tokyo~\cite{CosmoGrid}. The supercomputers
were interconnected by a lightpath with 10 Gigabit/s bandwidth capacity.  Our
main simulation consisted of $2048^3$ particles, and required about 10\% of its
runtime to exchange data over the wide area network. 

\begin{figure}
  \centering
  \includegraphics[width=0.95\textwidth]{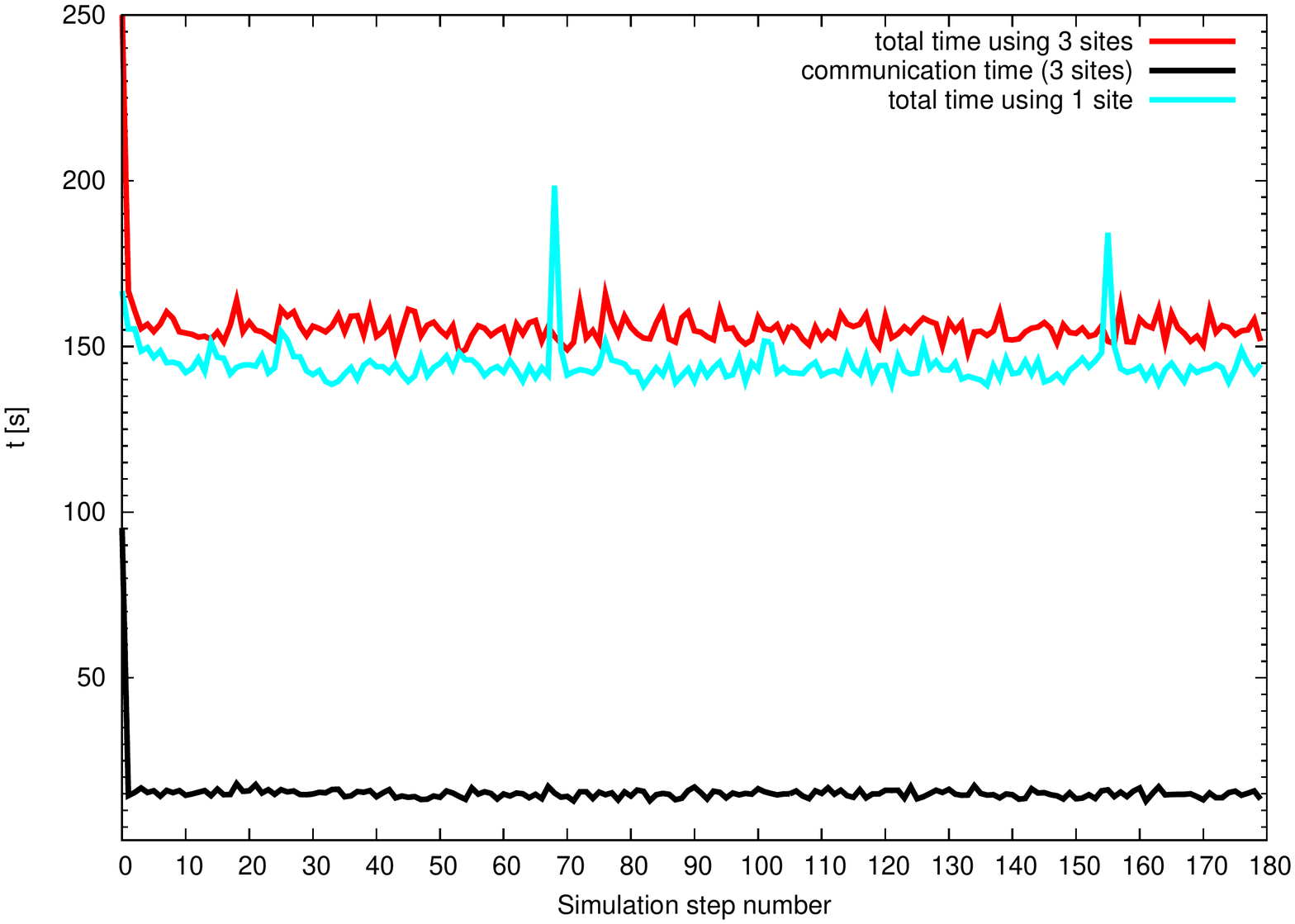}

  \caption{Comparison of the wallclock time required per simulation step
between a run using 2048 cores on one supercomputer (given by the teal line),
and a nearly identical run using 2048 cores distributed over three
supercomputers (given by the red line). The two peaks in the performance of the
single site run were caused by the writing of 160GB snapshots during those
iterations. The run over three sites used MPWide to pass data between
supercomputers. The communication overhead of the run over three sites is given
by the black line. See Groen et al.~\cite{Groen:2011-3} for a detailed
discussion on these performance measurements.}\label{fig:gbbp-run}

\end{figure}

\begin{figure}
  \centering
  \includegraphics[width=0.5\textwidth]{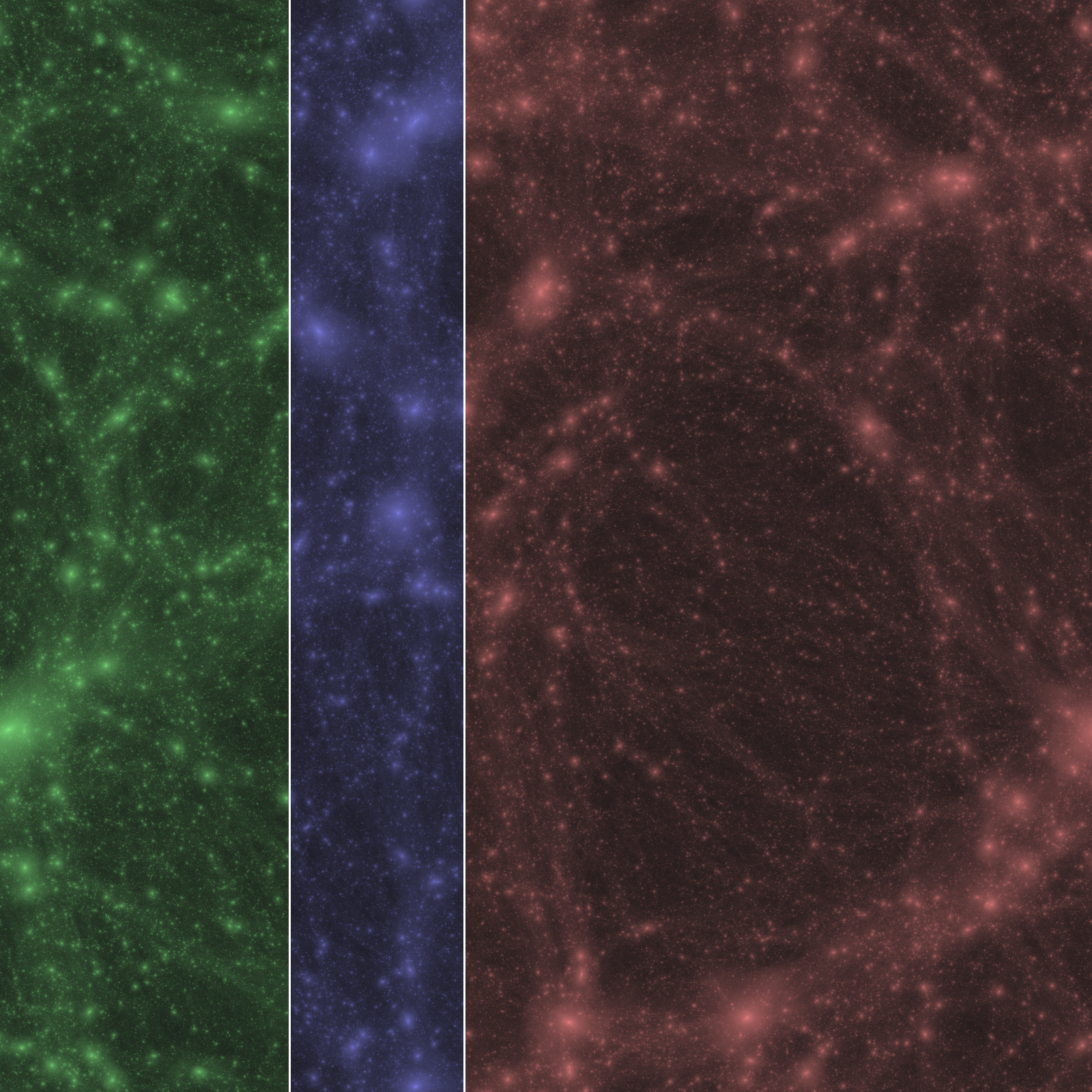}

  \caption{Snapshot of the cosmological simulation discussed in
Fig.~\ref{fig:gbbp-run}, taken at redshift $z$ = 0 (present day). The contents
have been colored to match the particles residing on supercomputers in Espoo
(green, left), Edinburgh (blue, center) and Amsterdam (red, right)
respectively~\cite{Groen:2011-3}.}\label{fig:gbbp-decomp}

\end{figure}

We subsequently extended the GreeM code, and used MPWide to run cosmological
simulations in parallel across up to 4 supercomputers~\cite{sushi}.  We also
performed a distributed simulation across 3 supercomputers, which consisted of
$2048^3$ particles and used 2048 cores in total~\cite{Groen:2011-3}. These
machines were located in Espoo (Finland), Edinburgh (Scotland) and Amsterdam
(the Netherlands). The run used MPWide version 1.0 and lasted for about 8 hours
in total. We present some performance results of this run in
Fig.~\ref{fig:gbbp-run}, and also provide the performance of the simulation
using one supercomputer as a reference. The distributed simulation is only 9\%
slower than the one executing on a single site, even though simulation data is
exchanged over a baseline of more than 1500 kilometres at every time step. A
snapshot of our distributed simulation, which also features dynamic load
balancing, can be found in Fig.~\ref{fig:gbbp-decomp}. The results from the
CosmoGrid project have been used for the analysis of dark matter
haloes~\cite{cosmogridscience} as well as for the study of star clusters in a
cosmological dark matter environment~\cite{Rieder:2013,Rieder:2013-2}.

\subsubsection{Distributed multiscale modelling of bloodflow}

\begin{figure}
  \centering
  \includegraphics[width=0.6\textwidth]{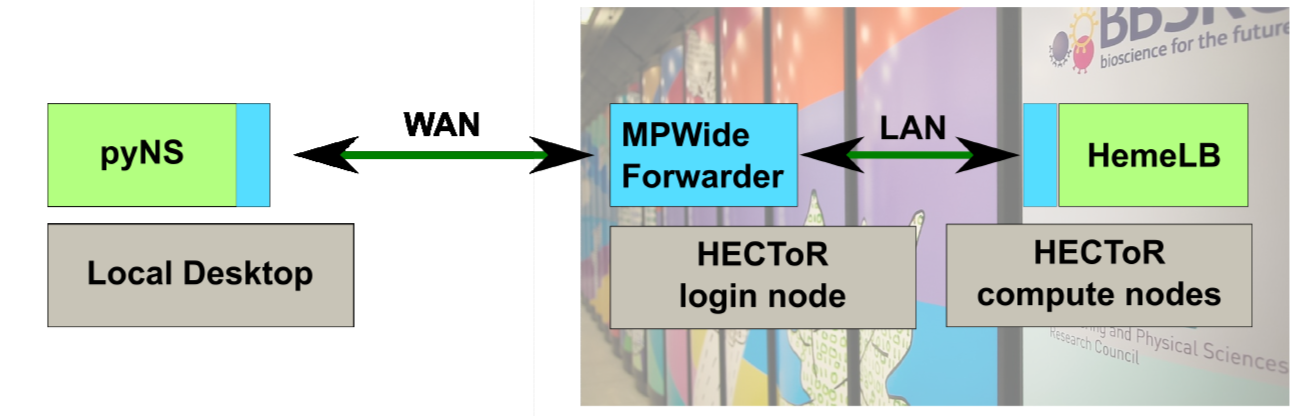}

  \caption{Overview of the codes and communication processes in the distributed 
multiscale bloodflow simulation. Here the 1D pyNS code uses MPWide to connect 
to an MPWide data forwarding process on the front-end node of the HECToR 
supercomputer. The 3D HemeLB code, which is executed on the compute nodes of
the HECToR machine also connects to this data forwarding process. The forwarding
process allows us to construct this simulation, even when the incoming ports of 
HECToR are blocked, and when the nodes where HemeLB will run are not known 
in advance. Once the connections are established, the simulations startd and 
boundary data is exchanged between the codes at runtime.}\label{fig:hemelb-pyns}

\end{figure}

We have also used MPWide to couple a three-dimensional cerebral bloodflow
simulation code to a one-dimensional discontinuous Galerkin solver for
bloodflow in the rest of the human body~\cite{Groen:2012-2}. Here, we used 
the 1D model to provide more realistic flow boundary conditions to the 3D model,
and relied on MPWide to rapidly deliver updates in the boundary conditions
between the two codes. We
ran the combined multiscale application on a distributed infrastructure, using
2048 cores on the HECToR supercomputer to model the cerebral bloodflow
and a local desktop at University College London to model the bloodflow in the
rest of the human body. The two resources are connected by regular internet,
and messages require 11 ms to traverse the network back and forth between the desktop and the 
supercomputer. We provide an overview of the technical layout of the codes and the 
communication processes in Fig.~\ref{fig:hemelb-pyns}

The communications between these codes are particularly frequent, as the codes
exchanged data every 0.6 seconds. However, due to latency hiding techniques we
achieve to run our distributed simulations with neglishible coupling overhead
(6 ms per coupling exchange, which constituted 1.2\% of the total runtime). A
full description of this run is provided by Groen et al.~\cite{Groen:2012-2}.

\subsubsection{Other research and technical projects}\label{Sec:OtherResearch}

We have used MPWide for several other purposes. First, MPWide is part of the
MAPPER software infrastructure~\cite{Zasada:2012}, and is integrated in the
MUSCLE2 coupling environment~\footnote{MUSCLE2 -
http://www.qoscosgrid.org/trac/muscle}. Within MUSCLE2, MPWide is used to
improve the wide area communication performance in coupled distributed
multiscale simulation~\cite{Borgdorff:2013}. Additionally, we 
applied the mpw-cp file transferring tool to test the network performance
between the campuses of University College London and Yale University.  In
these throughput performance tests we were able to exchange 256 MB of data at a
rate of $\sim$8 MB/s using {\tt scp}, a rate of $\sim$40 MB/s using MPWide, and
a rate of $\sim$48 MB/s using a commercial, closed-source file transfer tool
named Aspera.

We have conducted a number of basic performance tests over regular internet,
comparing the performance of MPWide with that of ZeroMQ~\footnote{ZeroMQ -
http://www.zeromq.org}, MUSCLE 1 and regular {\tt scp}.  During each test we
exchanged 64MB of data (in memory in the case of MPWide, MUSCLE and ZeroMQ, and from
file in the case of {\tt scp}), measuring the time to completion at least 20 times in
each direction. We then took the average value of these communications in each
direction. In these tests we used ZeroMQ with the default autotuned settings.

\begin{table}[!th]
    \centering
    \begin{tabular}{llll}

Endpoint 1  & Endpoint 2 & Name of tool & average throughput in each direction\\
            &            &              & MB $s^{-1}$\\
\hline
London, UK  & Poznan, PL & {\tt scp}    & 11/16 \\
London, UK  & Poznan, PL & MPWide       & 70/70 \\
London, UK  & Poznan, PL & ZeroMQ       & 30/110 \\
Poznan, PL  & Gdansk, PL & {\tt scp}    & 13/21 \\
Poznan, PL  & Gdansk, PL & MPWide       & 115/115 \\
Poznan, PL  & Gdansk, PL & ZeroMQ       & 64/- \\
Poznan, PL  & Amsterdam, NL & {\tt scp} & 32/9.1 \\
Poznan, PL  & Amsterdam, NL & MPWide    & 55/55 \\
Poznan, PL  & Amsterdam, NL & MUSCLE 1  & 18/18 \\

      \end{tabular}

  \caption{Summary of the throughput performance tests using MPWide and several
other tools to exchange data between resources in the United Kingdom (UK), the
Netherlands (NL) and Poland (PL) using regular internet. Tests over individual
connections were performed in quick succession to mitigate potential bias due
to background load on the internet backbone. A full report on these tests can
be found at http://www.mapper-project.eu, Deliverable 4.2 version 0.7.}

  \label{tab:mpw-performance}
\end{table}


\subsection{Implementation and architecture}

\begin{figure}
  \centering
  \includegraphics[width=0.9\textwidth]{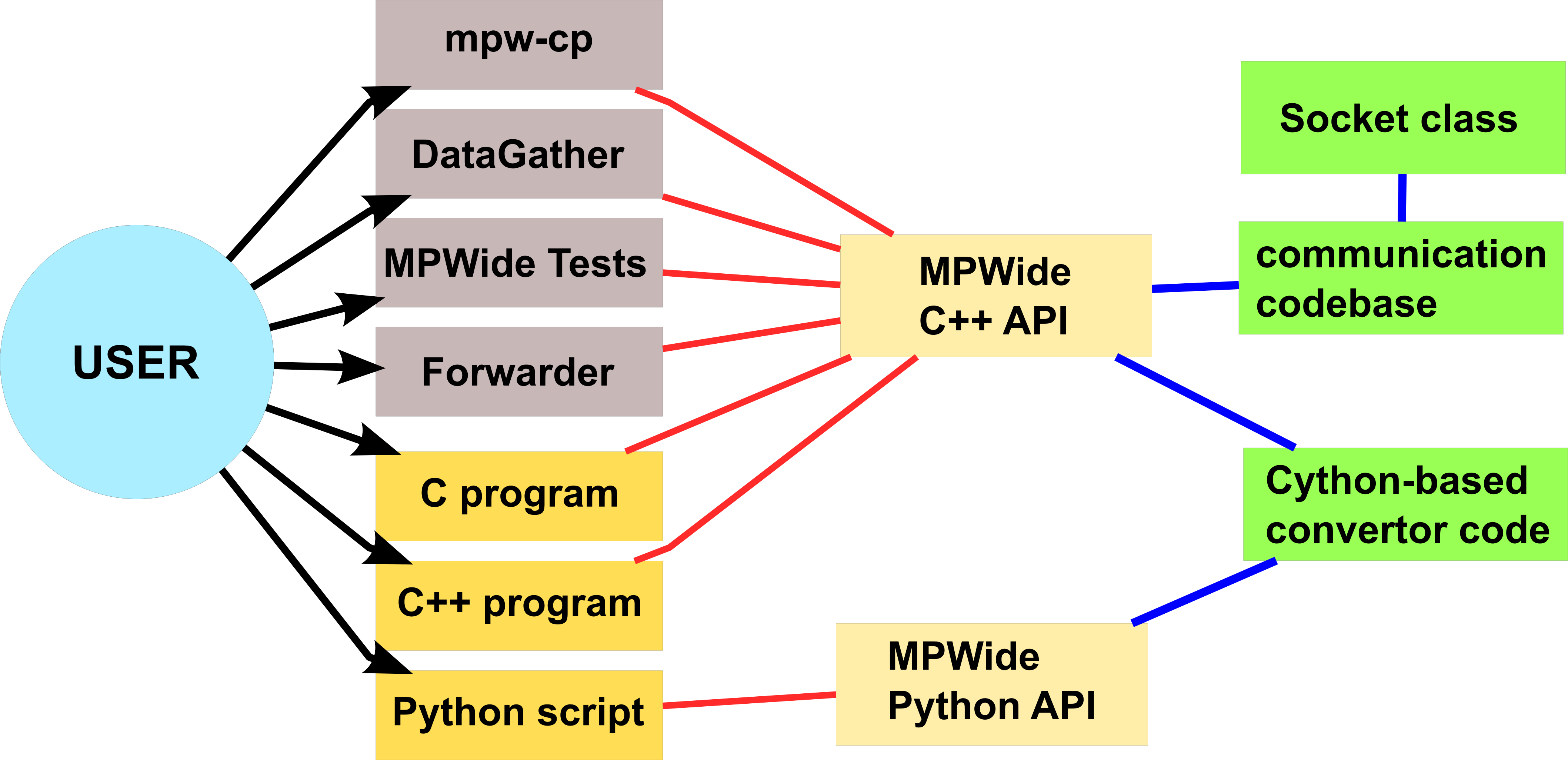}

  \caption{Overview of MPWide functionalities and their links to underlying
components. Functionalities available to the user are given by black arrows,
links of these functionalities to the corresponding MPWide API by red lines,
and internal codebase dependencies by dark blue
lines.}\label{fig:mpwide-design}

\end{figure}

We present a basic overview of the MPWide architecture in Fig.~\ref{fig:mpwide-design}.
MPWide has been implemented with a strong emphasis on minimalism, relying on a small
and flexible codebase which is used for a range of functionalities. 

\subsubsection{Core MPWide library}

The core MPWide functionalities are provided by the MPWide C++ API, the
communication codebase, and the Socket class. Together, these classes comprise
about 2000 lines of C++ code. The Socket class is used to manage and use
individual {\tt tcp} connections, while the role of the communication codebase is to
provide the MPWide API functionalities in C++, using the Socket class.
We provide a short listing of functions in the C++ API in 
Table~\ref{tab:MPW-functions}. More complete information can be found in the 
MPWide manual, which resides in the {\tt /doc} subdirectory of the source code
tree.

MPWide relies on a number of data structures, which are used to make it easier
to manage the customized connections between endpoints. The most
straightforward way to construct a connection in MPWide is to create a
communication {\em path}. Each path consists of 1 or more {\tt tcp} {\em
streams}, each of which is used to facilitate actual communications over that
path. Using a single {\tt tcp} stream is sufficient to enable a connection, but
in many wide area networks, MPWide will deliver much better performance when
multiple streams are used. MPWide supports the presence of multiple paths, and
the creation and deletion of paths at runtime. In addition, any messages can be
passed from one path to another using {\tt MPW\_Cycle()}, or {\tt MPW\_Relay()}
for sustained dedicated data forwarding processes (See Tab.~\ref{tab:MPW-functions}).

MPWide comes with a number of parameters which allow users to optimize the
performance of individual paths.  Aside from varying the number of streams,
users can modify the size of data sent and received per low-level communication
call (the {\em chunk size}), the {\tt tcp} window size, and limit the
throughput for individual streams by adjusting the {\em communication pacing
rate}. The number of streams will always need to provided by the user when
creating a path, but users can choose to have the other parameters
automatically tuned by enabling the MPWide autotuner.  The autotuner, which is
enabled by default, is useful for obtaining fairly good performance with
minimal effort, but the best performance is obtained by testing different
parameters by hand. When choosing the number of {\tt tcp} streams to use in a
path, we recommend using a single stream for connections between local
programs, and at least 32 streams when connecting programs over long-distance
networks. We have found that MPWide can communicating efficiently over as many
as 256 {\tt tcp} streams in a single path.

\begin{table}[!th]
    \centering
    \begin{tabular}{|l|l|}

\hline
function name  & summary description \\
\hline
{\tt MPW\_Barrier()}     & Synchronize between two ends of the network.             \\
{\tt MPW\_CreatePath()}  & Create and open a path consisting of 1+ {\tt tcp} streams. \\
{\tt MPW\_Cycle()}       & Send buffer over one set of channels, receive from other.\\
{\tt MPW\_DCycle()}      & As Cycle(), but with buffers of unknown size using caching.\\
{\tt MPW\_DestroyPath()} & Close and destroy a path consisting of 1+ {\tt tcp} streams.\\
{\tt MPW\_DNSResolve()}  & Obtain an IP address locally, given a hostname.          \\
{\tt MPW\_DSendRecv()}   & Send/receive buffers of unknown size using caching.      \\
{\tt MPW\_Init()}        & Initialize MPWide.                                       \\
{\tt MPW\_Finalize()}    & Close connections and delete MPWide buffers.             \\
{\tt MPW\_Recv()}        & Receive a single buffer (merging the incoming data).     \\
{\tt MPW\_Relay()}       & Forward all traffic between two channels.                \\
{\tt MPW\_Send()}        & Send a single buffer (splitted evenly over the channels).\\
{\tt MPW\_SendRecv()}    & Send/receive a single buffer.                            \\
\hline
{\tt MPW\_ISendRecv()}   & Send and/or receive data in a non-blocking mode.         \\
{\tt MPW\_Has\_NBE\_Finished()}& Check if a particular non-blocking call has completed.\\
{\tt MPW\_Wait()}        & Wait until a particular non-blocking call has completed.\\
\hline
{\tt MPW\_setAutoTuning()} & Enable or disable autotuning (default: enabled)        \\
{\tt MPW\_setChunkSize()}  & Change the size of data sent and received per low-level {\tt tcp} send command.\\
{\tt MPW\_setPacingRate()} & Adjust the software-based communication pacing rate.\\
{\tt MPW\_setWin()}        & Adjust the TCP window size within the constraints of the site configuration.\\
\hline
      \end{tabular}
  \caption{List of available functions in the MPWide API.}
  \label{tab:MPW-functions}
\end{table}

\subsubsection{Python extensions}

We have constructed a Python interface, allowing MPWide to be used through
Python~\footnote{Python - http://www.python.org}. We construct the interface
using Cython~\footnote{Cython - http://www.cython.org}, so as a result a recent
version of Cython is recommended to allow a smooth translation. The interface
works similar to the C++ interface, but supports only a subset of the MPWide 
features. It also includes a Python test script. We also implemented an 
interface using SWIG, but recommend Cython over SWIG as it is more portable. 

\subsubsection{Forwarder}

It is not uncommon for supercomputing infrastructures to deny direct
connections from the outside world to compute nodes. In privately owned
infrastructures, administrators commonly modify firewall rules to facilitate
direct data forwarding from outside to the compute nodes. The {\tt Forwarder} is a
small program that mimicks this behavior, but is started and run by the user,
without the need for administrative privileges. Because the {\tt Forwarder} operates
on a higher level in the network architecture, it is generally slightly less
efficient than conventional firewall-based forwarding. An extensive example of
using multiple {\tt Forwarder} instances in complex networks of supercomputers can be
found in Groen et al.~\cite{sushi}

\subsubsection{mpw-cp}

mpw-cp is a command-line file transfer tool which relies on SSH. Its
functionality is basic, as it essentially uses SSH to start a file transfer
process remotely, and then links that process to a locally executed one. mpw-cp
works largely similar to scp, but provides superior performance in many cases,
allowing users to tune their connections (e.g., by using multiple streams)
using command-line arguments. 

\subsubsection{DataGather}

The DataGather is a small program that allows users to keep two directories
synchronized on remote machines in real-time. It synchronizes in one direction
only, and it has been used to ensure that the data generated by a distributed
simulation is collected on a single computational resource. The DataGather can
be used concurrently with other MPWide-based tools, allowing users to
synchronize data while the simulation takes place.

\subsubsection{Constraints in the implementation and architecture}

MPWide has a number of constraints on its use due to the choices we made during
design and implementation. First, MPWide has been developed to use
the {\tt tcp} protocol, and is not able to establish or facilitate messages
using other transfer protocols (e.g. UDP). Second, compared to most MPI
implementations, MPWide has a limited performance benefit (and sometimes even a
performance disadvantage) on local network communications. This is because
vendor MPI implementations tend to contain architecture-specific optimizations
which are not in MPWide. 

Third, MPWide does not support explicit data types in
its message passing, and treats all data as an array of characters. We made this 
simplification, because data types vary between different
architectures and programming environments. Incorporating the management of these
in MPWide would result in a vast increase of the code base, as well as a permanent 
support requirement to update the type conversions in MPWide, whenever a new 
platform emerges. We recommend that users perform this serialization task in their
applications, with manual code for simple data types, and relying on a high-quality 
serialization libraries for more complex data types. 

\subsection{Quality Control}

Due to the small size of the codebase and the development team, MPWide has a 
rather simplistic quality control regime. Prior to each public release, the 
various functionalities of MPWide are tested manually for stability and 
performance. Several test scripts (those which do not involve the use of 
external codes) are available as part of the MPWide source distribution, 
allowing users to test the individual functionalities of MPWide without 
writing any new code of their own. These include:
\begin{itemize}
\item MPWUnitTests - A set of basic unit tests, can be run without any additional arguments.
\item MPWTestConcurrent - A set of basic functional tests, can be run without any additional arguments.
\item MPWTest - A benchmark suite which requires to be started manually on both end points.
\end{itemize}

More details on how to use these tests can be found in the manual, which is 
supplied with MPWide.

\section{Availability}

\subsection{Operating system}

MPWide is suitable for most Unix environments. It can be installed and used as-is 
on various supercomputer platforms and Linux distributions. We have also been able
to install and use this version of MPWide successfully on Mac OS X.

\subsection{Programming language}

MPWide requires a C++ compiler with support for pthreads and UNIX sockets.

\subsection{Additional system requirements}

MPWide has no inherent hardware requirements.

\subsection{Dependencies}

MPWide itself has no major dependencies. The mpw-cp functionality relies on
{\tt SSH} and the Python interface has been tested with Python 2.6 and 2.7. The
Python interface has been created using SWIG, which is required to generate
a new interface for different types of Python, or for non 64-bit and/or non-Linux
platforms.

\subsection{List of contributors}

\begin{itemize}
\item Derek Groen, has written most of MPWide and is the main contributor to this writeup.
\item Steven Rieder, assisted in testing MPWide, provided advice during development, and contributed to the writeup.
\item Simon Portegies Zwart, provided supervision and support in the MPWide development, and contributed to the writeup.
\item Joris Borgdorff, provided advice on the recent enhancements of MPWide, and made several recent contributions to the codebase.
\item Cees de Laat, provided advice during development and helped arrange the initial Amsterdam-Tokyo lightpath for testing and production.
\item Paola Grosso, provided advice during development and in the initial writeup of MPWide.
\item Tomoaki Ishiyama, contributed in the testing of MPWide and implemented the first MPWide-enabled application (the GreeM N-body code).
\item Hans Blom, provided advice during development and conducted preliminary tests to compare a TCP-based with a UDP-based approach.
\item Kei Hiraki, provided advice during development and infrastructural support during the initial wide area testing of MPWide.
\item Keigo Nitadori, provided advice during development.
\item Junichiro Makino, for providing advice during development.
\item Stephen L.W. McMillan, provided advice during development.
\item Mary Inaba, provided infrastructural support during the initial wide area testing of MPWide.
\item Peter Coveney, provided support on the recent enhancements of MPWide.
\end{itemize}


\subsection{Software location}

We have made MPWide available on GitHub at: https://github.com/djgroen/MPWide.

\subsection{Code Archive}
Name: MPWide version 1.8.1
Persistent identifier: http://dx.doi.org/10.6084/m9.figshare.866803
Licence: MPWide has been released under the Lesser GNU Public License version 3.0.
Publisher: Derek Groen
Date published: 3rd of December 2013

\subsection{Code Repository} 
Name: MPWide
Identifier: https://github.com/djgroen/MPWide
Licence: MPWide has been released under the Lesser GNU Public License version 3.0.
Publisher: Derek Groen (account name: djgroen)
Date published: 15th of October 2013.

\subsection{Language}
GitHub uses the git repository system. The full MPWide distribution contains code written
primarily in C++, but also contains fragments written in C and Python.

The code has been commented and documented solely in English.

\section{Reuse potential}

MPWide has been designed with a strong emphasis on reusability. It has a small
codebase, with minimal dependencies and does not make use of the more obscure
C++ features. As a result, users will find that MPWide is trivial to set up in
most Unix-based environments. MPWide does not receive any official funding for
its sustainability, but the main developer (Derek Groen) is able to respond to 
any queries and provide basic assistance in adapting MPWide for new applications.

\subsection{Reuse of MPWide}

MPWide can be reused for a range of different purposes, which all share one 
commonality: the combination of light-weight software with low latency and 
high throughput communication performance. 

MPWide can be reused to parallelize an application across
supercomputers and to couple different applications running on different
machines to form a distributed multiscale simulation. A major advantage of
using MPWide over regular TCP is the more easy-to-use API (users do not have to
cope with creating arrays of sockets, or learn low-level TCP calls such as
listen() and accept()), and built-in optimizations that deliver superior
performance over long-distance networks. 

In addition, users can apply MPWide to facilitate high speed file transfers
over wide area networks (using {\tt mpw-cp} or the DataGather). MPWide provides
superior performance to existing open-source solutions on many long-distance
networks (see e.g., section~\ref{Sec:OtherResearch}). MPWide could also be
reused to stream visualization data from an application to a visualization
facility over long-distances, especially in the case when dedicated light paths
are not available.

Users can also use MPWide to link a Python program directly to a C
or C++ program, providing a fast and light-weight connection between different 
programming languages. However, the task of converting between data types is 
left to the user (MPWide works with character buffers on the C++ side, and 
strings on the Python side).

\subsection{Support mechanisms for MPWide}

MPWide is not part of any officially funded project, and as such does not
receive sustained official funding. However, there are two mechanisms for
unofficial support.  When users or developers run into problems we encourage
them to either raise an issue on the GitHub page or, if urgent, to contact the
main developer (Derek Groen, djgroennl@gmail.com) directly. 

\subsection{Possibilities of contributing to MPWide}

MPWide is largely intended as stand-alone and a very light-weight communication
library, which is easy to maintain and support. To make this possible, we aim to
retain a very small codebase, a limited set of features, and a minimal number of
dependencies in the main distribution.  

As such, we are fairly strict in accepting new features and
contributions to the code on the central GitHub repository. We primarily aim to
improve the performance and reliability of MPWide, and tend to accept new contributions
to the main repository only when these contributions boost these aspects of the
library, and come with a limited code and dependency footprint.

However, developers and users alike are free to branch MPWide into a separate
repository, or to incorporate MPWide into higher level tools and services, as
allowed by the LGPL 3.0 license. We strongly recommend integrating MPWide as a
library module directly into higher level services, which then rely on the
MPWide API for any required functionalities. MPWide has a very small code
footprint, and we aim to minimize any changes in the API between versions,
allowing these high-level services to easily swap their existing MPWide module
for a future updated version of the library. We have already used this approach
in codes such as SUSHI, HemeLB and MUSCLE 2.

\section*{Funding statement}

This research is supported by the Netherlands organization for Scientific
research (NWO) grants \#614.061.608 (AMUSE), \#614.061.009 (LGM), \#639.073.803,
\#643.000.803 and \#643.200.503, the European Commission grant for the QosCosGrid project
(grant number: FP6-2005-IST-5 033883), the Qatar National Research
Fund (QNRF grant code NPRP 5-792-2-328) and the MAPPER project (grant number:
RI-261507), SURFNet with the GigaPort project, NAOJ, the International
Information Science Foundation (IISF), the Netherlands Advanced School for
Astronomy (NOVA), the Leids Kerkhoven-Bosscha fonds (LKBF) and the Stichting
Nationale Computerfaciliteiten (NCF). SR acknowledges support by the John 
Templeton Foundation, grant nr. FP05136-O. We thank the organizers of the Lorentz
Center workshop on Multiscale Modelling and Computing 2013 for their support.
We also thank the DEISA Consortium (www.deisa.eu), co-funded through the EU FP6
project RI-031513 and the FP7 project RI-222919, for support within the DEISA
Extreme Computing Initiative (GBBP project).

\bibliographystyle{plain}
\bibliography{Library}

\end{document}